\documentclass[10pt]{article}
\usepackage{amsfonts}
\usepackage{color,caption}
\usepackage{graphicx}
\usepackage{amsmath}
\usepackage{float}
\usepackage{amssymb}
\providecommand{\U}[1]{\protect\rule{.1in}{.1in}}
\textwidth=6.5in \hoffset=-.75in \voffset=-1in \numberwithin{equation}{section}
\numberwithin{figure}{section}

\newcommand {\be}{\begin{equation}}
 \newcommand {\ee}{\end{equation}}
 \newcommand {\bea}{\begin{eqnarray}}
 
 \newcommand {\eea}{\end{eqnarray}}

\begin{document}
\begin{titlepage}
\bigskip \begin{flushright}
\end{flushright}
\vspace{1cm}
\begin{center}
{\Large \bf {Dual CFT on Dyonic Kerr-Sen black hole and its gauged and ultraspinning counterparts}}\\
\vskip 1cm
\end{center}
\vspace{1cm}
\begin{center}
Muhammad F. A. R. Sakti{\footnote{fitrahalfian@gmail.com}} and Piyabut Burikham{\footnote{piyabut@gmail.com}},
\\
High Energy Physics Theory Group, Department of Physics, Faculty of Science, Chulalongkorn University, Bangkok 10330, Thailand \\

\vspace{1cm}
\end{center}

\begin{abstract}
We demonstrate a strong evidence of entropy matching that rotating dyonic black holes in Einstein-Maxwell-Dilaton-Axion (EMDA) theory is holographically dual to a 2D conformal field theory. We first investigate the duality on dyonic Kerr-Sen black hole with non-vanishing dilaton and axion charges. The near-horizon geometry of extremal dyonic Kerr-Sen spacetime possesses the $SL(2,R) \times U (1)$ isometry where the asymptotic symmetry group method can be used to find the corresponding central charge. We find two different branches of masses which correspond to CFT with two different central charges $c_L=12am_+$ and $c_L=12am_-$. The exact agreement between the Bekenstain-Hawking entropy and entropy from CFT is then found also in two different branches of extremal entropy. Furthermore, we demonstrate that this duality is robust insofar for non-zero AdS length. The duality holds for both dyonic Kerr-Sen-AdS black hole and its ultraspinning counterpart. In both cases, we obtain the expected entropy from CFT which matches exactly with the Bekenstein-Hawking entropy. Since dyonic and axion charges are proportional to $1/m$, we note that there are possibly more than two branches of the central charge for non-zero AdS length in terms of mass. When we turn off dyonic charge, the axion charge vanishes giving the results of Kerr-Sen-AdS black hole. Moreover, when we assume the equal electromagnetic charges, it recovers the results when the dilaton charge vanishes. Lastly, we compare the results of dyonic Kerr-Sen-AdS black hole and its ultraspinning counterpart to those of the dyonic Kerr-Newman-AdS black hole and the ultraspinning counterpart. Depending on the dyonic charge parameters, it is found that extremal ultraspinning dyonic Kerr-Sen-AdS black hole is not always superentropic.
\end{abstract}
\end{titlepage}\onecolumn
\bigskip

\section{Introduction}

It has been pointed out by Bekenstein and Hawking that black holes must have an entropy to prevent the violation of second law of thermodynamics. It also corresponds to the existence of radiation from the black holes with finite temperature. These thermodynamic behaviors of black holes indicate that there exists the underlying micro structure within the black holes. One big question then arises from this theory questioning the explanation of the origin of the black hole microstates. It has been a main curiosity for decades to explain the Bekenstein-Hawking entropy from this microscopic point of view. Nonetheless, there is yet no complete answer to this question. However, for a large class of extremal supersymmetric black holes, this question has been answered in which the Bekenstein-Hawking entropy can be computed also by counting the degeneracy of Bogomol’nyi-Prasad-Sommerfield soliton bound states \cite{StromingerVafaPLB1996,SenGRG2008}. 

Another answer to the previous question is the Kerr/CFT duality. This duality states that there is a correspondence between the associated physical quantities of the extremal four-dimensional Kerr black hole and almost similar physical quantities in a chiral conformal field theory (CFT) \cite{GuicaPRD2009}. This statement is parallel to the general one given by Brown and Hanneaux for AdS$_3$ \cite{brown1986}. However, the AdS$_3$ spacetime is replaced by near-horizon extremal black hole metric. In the near-horizon region of extremal black holes, it is shown that there is a conformal invariance implying the presence of an AdS structure on the spacetime with $SL(2,R)\times U(1)$ isometry. The generalization of this correspondence to the other extremal rotating black holes, in four and higher dimensions, and beyond Einstein theory have been extensively studied during the last decade \cite{Hartman2009, Ghezelbash2009, Lu2009, Li2010, Anninos2010, Ghodsi2010, Ghezelbash2012, Astorino2015, Astorino2015a, Siahaan2016, Astorino2016, Sinamuli2016, Sakti2018, SaktiEPJPlus2019, SaktiMicroJPCS2019,SaktiAnnPhys2020,SaktiPhysDarkU2021}. The 2D chiral CFT has been carried out by considering the asymptotic symmetry group with some proposed boundary conditions, from which we can generate a class of diffeomorphisms of the near-horizon Kerr geometry.  After defining charges associated with the diffeomorphisms and evaluating the Dirac brackets of the charges, we can produce a Virasoro algebra with non-vanishing (quantum) central charge which is the extension of $U(1)$ isometry of the near-horizon extremal black hole geometry. The entropy is then calculated using the Cardy's growth of states for a 2D CFT which is the function of central charge and temperature. The temperature can then be calculated by assuming Frolov-Thorne vacuum. From this duality calculation, the exact matching between entropy from CFT and Bekenstein-Hawking entropy of extremal Kerr black hole is established. 

Interestingly, the Kerr/CFT correspondence may apply not only to the extremal black holes, but also to non-extremal black ones. The hidden conformal symmetry of the Kerr black hole has been revealed for the first time in Ref.~\cite{Castro2010}. The conformal symmetry appears on the solution space of the scalar probe in the black hole background. Further studies show that not only scalar probe, yet higher spin particles can also be the probe to reveal the hidden conformal symmetry \cite{LoweSkanataPRD2014}. Similar to the Kerr/CFT in extremal black holes, this also can be extended to various black hole solutions in four and five dimensions, and also in different gravitational theories \cite{ChenLong2010, CveticLarsenJHEP2009, ChenSun2010, ChenWang2010, WangLiu2010, ChenLongJHEP2010, SetareKamali2010, GhezelbashKamali2010, ChenHuangPRD2010, ChenChen2011, ChenGhezelbash2011, ChenZhangJHEP2011, HuangYuan2011, Shao2011, DeyouChen2011, CveticLarsenJHEP2012a, CveticLarsenJHEP2012b, GhezelbashSiahaanCQG2013, GhezelbashSiahaanGRG2014, SiahaanAcc2018, Saktideformed2019, SaktiNucPhysB2020,SaktiPhysDarkU2022,Saktidyonicnonex}. The revelation of the hidden conformal symmetry shows that in the near-horizon region and low-frequency limit, the phase space possesses $SL(2,R)\times SL(2,R)$ isometry similar to AdS$_3$ spacetime. In these papers, the central charges are computed using numerical observation in order to match the Bekenstein-Hawking entropy with the Cardy entropy. Another observation for non-extremal Kerr black hole is that we can identify two sets of Virasoro diffeomorphisms acting on the horizon leading to non-vanising left- and right-moving central charges \cite{HacoJHEP2018}. This remarkable calculation fills up the gap left from the first revelation of hidden conformal symmetry in the non-extremal black hole. Another intriguing result is provided in Ref.~\cite{AggarwalCastroJHEP2020} that one can also extend the conventional CFT to the warped CFT in order to calculate the entropy for non-extremal Kerr black holes. Therein the non-trivial diffeomorphisms lead to a Virasoro-Kac-Moody algebra with non-trivial central extensions. 

Inspired by the Kerr/CFT correspondence, it is natural to extend the similar calculation to the family of black holes in Einstein-Maxwell-Dilaton-Axion (EMDA) theory. We want to study the thermodynamic properties of dyonic Kerr-Sen black hole, its gauged, and ultraspinning counterparts. Non-dyonic Kerr-Sen black hole has been investigated in Ref.~\cite{Ghezelbash2009} for extremal case and in Ref.~\cite{GhezelbashSiahaanCQG2013} for non-extremal case showing that the central charge of this black hole is identical to Kerr black hole which is the solution to Einstein vacuum field equation, although Kerr-Sen also possesses non-vanishing electric and dilaton charges. The solution of dyonic case is provided in Ref.~\cite{WuWuWuYuPRD2021} which is the extension of Ref.~\cite{WuWuYuWuPRD2020} for non-dyonic case, yet with non-zero cosmological constant (AdS length). This black hole family emerges in the low-energy heterotic string theory that includes dilaton and axion. Since there exist scalar fields representing dilaton and axion, this solution will differ from dyonic Kerr-Newman~(-AdS) black hole solution. 

One fascinating limit of gauged dyonic black hole is the ultraspinning limit, i.e. the spin of the black hole is boosted to the value of AdS length. In the ultraspinning limit, the Kerr-Newman-AdS black hole solution possesses non-compact horizon and becomes superentropic. The non-compact horizon occurs due to the presence of conical singularity while superentropic means that for a given volume $V$, the entropy will be maximized \cite{HennigarMannPRL2015}. So far, the study for Kerr-Newman-AdS black hole has been done in Ref.~\cite{Sinamuli2016} in the context of Kerr/CFT correspondence. In the present paper, it will be demonstrated that the Kerr/CFT correspondence is also valid for dyonic black hole family in EMDA theory and indeed robust. First, we consider dyonic Kerr-Sen black hole and show that there are similar features between dyonic and non-dyonic cases. This work is then extended to the gauged counterpart, then in the situation where the cosmological constant is non-zero, and finally when the ultraspinning limit is reached. It is intriguing also to compare the results with the dyonic Kerr-Newman-AdS black hole family.

The paper is organized as follows.  After the introduction, we briefly review the dyonic Kerr-Sen black hole and its thermodynamic properties in Section~\ref{DKSBH}. In Section~\ref{sec:KerrCFT}, we investigate the near-horizon extremal dyonic Kerr-Sen black hole and derive its entropy using the Kerr/CFT correspondence. In Section~\ref{sec:KerrCFTgauge} and \ref{sec:KerrCFTultra}, we extend the work from previous section to the gauged dyonic Kerr-Sen-AdS black hole and its ultraspinning counterpart, respectively. In Section~\ref{DKSKNCom}, we provide the comparison between the dual CFT of dyonic Kerr-Sen-AdS and dyonic Kerr-Newman-AdS black holes. Section~\ref{eq:conclusion} concludes our work.

\section{Dyonic Kerr-Sen black holes}  \label{DKSBH}

The dyonic NUT generalization of Kerr-Sen black hole solution was first given in Ref.~\cite{GaltsovPRL1995} in ungauged case. \textcolor{black}{However, in this paper, we set the NUT parameter to zero, and consider only the dyonic Kerr-Sen black hole. The dyonic generalization of Kerr-Sen solution without NUT parameter} is re-written in Ref.~\cite{WuWuWuYuPRD2021} in a simpler form using certain coordinate transformations. In Ref.~\cite{WuWuWuYuPRD2021}, the gauged solution of EMDA theory is also provided in a simpler way coming from the original derivation in Refs.~\cite{ChongCveticNPB2005,ChowComperePRD2014}. In this section, we will briefly review the spacetime metric of the dyonic Kerr-Sen black holes which is given in Ref.~\cite{WuWuWuYuPRD2021}. 
The Lagrangian density of the EMDA theory that is considered to find dyonic Kerr-Sen black hole is given by \cite{ChowCompereCQG2014,ChowComperePRD2014}
\begin{equation}
\mathcal{L}=\sqrt{-g}\left[R - \frac{1}{2}(\partial\phi)^2-\frac{1}{2}e^{2\phi}(\partial\chi)^2-e^{-\phi}F^2\right] +\frac{\chi}{2}\epsilon^{\mu\nu\rho\lambda}F_{\mu\nu}F_{\rho_\lambda}, \label{eq:Lagrangian}\
\end{equation}
where $\epsilon^{\mu\nu\rho\lambda}$ is the Levi-Civita antisymmetric tensor density in 4D. The dual of the gauge potential $B$ is defined by $dB=-e^{-\phi}\star F - \chi F$. \textcolor{black}{The theory presented in Lagrangian density (\ref{eq:Lagrangian}) has an $SL(2,R)$ global symmetry that can transform the dyonic black hole to purely electric or magnetic black hole}. The dyonic Kerr-Sen black hole solution to that Lagrangian density is given in Eq. (4) in Ref.~\cite{WuWuWuYuPRD2021}.

However, as mentioned in Ref.~\cite{WuWuWuYuPRD2021}, the the dyonic Kerr-Sen black hole spacetime metric can also be written in a more symmetric way by shifting the radial coordinate as $r \rightarrow r+ d$. In this coordinate, the spacetime metric together with the electromagnetic potential, the dual electromagnetic potential, dilaton field, and axion field, are given by
\begin{eqnarray}
	ds^2 & = & - \frac{\Delta}{\varrho^2 }\hat{X}^2+  \frac{\varrho ^2}{\Delta}d\hat{r}^2 + \varrho ^2 d\theta^2 + \frac{\sin^2\theta}{\varrho ^2}\hat{Y}^2,\label{eq:dKSmetric} \\
	\textbf{A} &=& \frac{q(\hat{r}+d-p^2/m)}{\varrho^2}\hat{X}-\frac{p\cos\theta}{\varrho^2}\hat{Y}, ~~~	\textbf{B} = \frac{p(\hat{r}+d-p^2/m)}{\varrho^2}\hat{X}+\frac{q\cos\theta}{\varrho^2}\hat{Y}, \label{eq:BdKS}\\
	e^{\phi} &=& \frac{(\hat{r}+d)^2+(k+a\cos\theta)^2}{\varrho^2}, ~~~	\chi = 2\frac{k\hat{r}-da\cos\theta}{(\hat{r}+d)^2+(k+a\cos\theta)^2}, \label{eq:axiondKS}
\end{eqnarray}
respectively, where
\begin{equation}
	\hat{X}=d\hat{t} - a \sin^2\theta d\hat{\phi}, ~~~ 
	\hat{Y}=ad\hat{t}- (\hat{r}^2-d^2-k^2+a^2 ) d\hat{\phi},
\end{equation}
\begin{equation}
	\varrho^2 =\hat{r}^2-d^2 -k^2 + a^2\cos^2\theta , ~~~~ \Delta = \hat{r}^2 -2m \hat{r} -d^2 -k^2+a^2 +p^2 +q^2. \nonumber\
\end{equation}
We have to note that $m, a, q, p, d, k$ are mass, spin, electric charge, magnetic (dyonic) charge, dilaton charge, and axion charge of the black hole, respectively. The dilaton and axion charges clearly depend on the electric and magnetic charges by the following relations
\begin{equation}
	d= \frac{p^2 -q^2}{2m}, ~~~~~ k =\frac{pq}{m}.\
\end{equation}
One also can write
\begin{equation}
	d^2+k^2 = \left( \frac{p^2+q^2}{2m}\right)^{2}. \label{eq:relationdkpq}\
\end{equation}
This black hole possesses inner and outer horizons as given by
\begin{equation}
	r_\pm = m \pm \sqrt{ m^2 + d^2 + k^2 -a^2 - p^2 - q^2}. \label{eq:rpmdKS}
\end{equation}

The dyonic Kerr-Sen black hole satisfies the following thermodynamic relation
\begin{eqnarray}
	dM= T_H dS+\Omega_H dJ +\Phi_H dQ +\Psi_H dP. \label{eq:thermoKS}
\end{eqnarray}
The quantities on above equation are given by
\begin{equation}
	M= m, ~~~ J = ma, ~~~ Q= q, ~~~ P= p, ~~~
	T_H = \frac{r_+ -m}{2\pi (r_+^2 -d^2-k^2 +a^2)}, ~~~
	S_{BH}=\pi(r_+^2 -d^2 -k^2 +a^2), \label{eq:SBHKS}
\end{equation}
\begin{equation}
	\Omega_H = \frac{a}{r_+^2 -d^2 -k^2 +a^2}, ~~~
	\Phi_H = \frac{q(r_+ +d -p^2/m)}{r_+^2 -d^2 -k^2 +a^2},~~~  \Psi_H = \frac{p(r_+ +d -p^2/m)}{r_+^2 -d^2 -k^2 +a^2}, \label{eq:potKS}
\end{equation}
where those are mass, angular momentum, electric charge, magnetic charge, Hawking temperature, Bekenstein-Hawking entropy, angular velocity, electric potential, and magnetic potential, respectively.

\textcolor{black}{Now we consider the extremal limit of dyonic Kerr-Sen black hole. In this limit, $m^2 + d^2 + k^2 =a^2 + p^2 + q^2$, and both horizons in Eq. (\ref{eq:rpmdKS}) coincide as $r_\pm=m$. Note that one can also write $d^2+k^2$ in terms of $q^2+p^2$ to reduce the number of parameters,
\begin{equation}	
m^{2} =a^2+q^2+p^2 -\left(\frac{q^2+p^2}{2m}\right)^{2}.
\end{equation}
Based on this relation, there are two branches of extremal dyonic Kerr-Sen black hole where the mass is given by
\begin{equation}
m^{2} = \frac{1}{2}(a^{2}+p^{2}+q^{2})\left[ 1\pm \sqrt{1-\left( \frac{p^{2}+q^{2}}{a^{2}+p^{2}+q^{2}}\right)^{2}} \right]. \label{mext}
\end{equation}
We identify each branch $m_{+(-)}$ with the plus~(minus) sign of the square root in (\ref{mext}) respectively. For this extremal black hole, the Hawking temperature in Eq. (\ref{eq:SBHKS}) vanishes while other thermodynamic quantities that contain mass term remain non-zero and possess two branches. These branches will reduce into one only when the spin approaches zero. In this case, the black hole solution will reduce to extremal dyonic GMGHS (Gibbons-Maeda-Garfinkle-Horowitz-Strominger) black hole with zero entropy~\cite{HerdeiroPLB2021}. This condition is distinctive from extremal (dyonic) Reissner-Nordstr{\"o}m solution which has non-zero entropy. Zero entropy is related to zero central charge in CFT. Vanishing $c_L$ is trivial and exactly denotes that the Virasoro algebra reduces to the classical Witt algebra. In the next section, we will study the thermodynamic properties of this extremal dyonic Kerr-Sen black hole using the Kerr/CFT correspondence.
}

\section{CFT Dual of Dyonic Kerr-Sen Black Hole}
\label{sec:KerrCFT}
The main upshot of this section is to provide the derivation of Cardy entropy \cite{GuicaPRD2009},
\begin{equation}
	S_{CFT}=\frac{\pi^2}{3}\left(c_L T_L +c_R T_R\right),\label{eq:Cardyentropy}
\end{equation}
for extremal black holes in EMDA theory given in this paper. We will derive the corresponding central charges $(c_L,c_R)$ using the asymptotic symmetry group (ASG) and the temperatures $(T_L,T_R)$ in order to compute the Cardy entropy of the black holes given in this paper. We apply the Cardy entropy formula for the black hole solutions in this paper for extremal case. First, in order to apply the asymptotic symmetry group calculation, we need to find the near-horizon geometry satisfying the $SL(2,R)\times U(1)$ isometry group. We will present explicitly the near-horizon geometry of the extremal rotating dyonic Kerr-Sen black hole. Then we will derive the non-trivial diffeomorphisms which are associated with non-vanishing conserved surface charges. It will be shown that the four-dimensional rotating dyonic Kerr-Sen black hole belongs to phase space representing one copy of the Virasoro algebra with a particular central charge. After finding the temperatures using the generalized Frolov-Thorne vacuum, one can compute the Cardy formula to match with the Bekenstein-Hawking entropy for black holes.

\subsection{Near-horizon extremal black hole metric}
\label{subsec:nearhorizonEMDA}
The near-horizon form of the spacetime metric can be obtained using the specific coordinate transformation representing near-horizon region approximation. First, we will find the near-horizon geometry of the extremal dyonic Kerr-Sen black hole. To find the near-horizon geometry of the extremal dyonic Kerr-Sen black hole (\ref{eq:dKSmetric}),  
we consider the following coordinate transformations \cite{Hartman2009,Compere2017}
\begin{eqnarray}
	\hat{r} = r_+ +\epsilon r_0 r,~~~ \hat{t} = \frac{r_0}{\epsilon} t, ~~~ \hat{\phi} = \phi + \frac{\Omega_H r_0}{\epsilon} t, \label{eq:extremaltransformation} \
\end{eqnarray}
where $r_0 $ is scaling constant where we may define as $r_0^2 = r_+^2 -d^2 -k^2 +a^2 $. In near-horizon limit, we have $ \epsilon \rightarrow 0 $. We can obtain the near-horizon extremal metric of Eq. (\ref{eq:dKSmetric}) which is given by
\begin{eqnarray}
	ds^2 = \Gamma(\theta)\left(-r^2 dt^2 + \frac{dr^2}{r^2} + d\theta ^2 \right)  +\gamma(\theta) \left(d\phi +e r dt\right)^2, \label{eq:extremalmetricnew}\
\end{eqnarray}
where the metric functions are given by
\begin{equation}
	\Gamma(\theta)= \varrho_+ ^2, ~~\gamma(\theta)=\frac{r_0^4 \sin^2\theta}{\varrho_+^2}, ~~
	\varrho_+^2 = m^2 -d^2 -k^2 + a^2\cos^2\theta, ~~ e = \frac{2am}{r_0^2}. \label{eq:metricfunctionnearhorizonKS}\nonumber\
\end{equation}
One can see that the AdS$_2$ factor emerges in this near-horizon metric denoting that the metric now has AdS$_2 \times S^2$ structure. This is the origin how we infer that the AdS/CFT correspondence for black holes, namely Kerr/CFT may apply.

Since there exists several non-vanishing scalar and vector fields, it is also applicable to use the near-horizon coordinate transformation to those fields. In the near-horizon limit, we find that the gauge field and its dual are given by
\begin{eqnarray}
	\textbf{A}_+ =f_a(\theta) \left(d\phi +e r dt \right)+\frac{q\left[(m +d)^2+k^2-a^2-2p^2 \right]}{e(m^2-d^2-k^2+a^2)}d\phi, \label{eq:nearhorizonelectromagneticpot1KS}\\
	\textbf{B}_+ =f_b(\theta) \left(d\phi +e r dt \right)+\frac{p\left[(m +d)^2+k^2-a^2-2p^2 \right]}{e(m^2-d^2-k^2+a^2)}d\phi,
	\label{eq:nearhorizonelectromagneticpotdual1KS}\
\end{eqnarray}
where the functions $f(\theta)$ for each field are given by
\begin{eqnarray}
	f_a(\theta) = \frac{q\left[2p^2-(m+ d)^2-k^2+a^2\cos^2\theta\right] +2apm\cos\theta}{2am \varrho_+^2}r_0^2, \label{eq:nearhorizonelectromagneticpot2}\\
	f_b(\theta) = \frac{p\left[2p^2-(m + d)^2-k^2+a^2\cos^2\theta\right] -2aqm\cos\theta}{2am \varrho_+^2}r_0^2,\
	\label{eq:nearhorizonelectromagneticpotdual2}
\end{eqnarray}
where we can gauge away the second term on the electromagnetic potential (\ref{eq:nearhorizonelectromagneticpot1KS}) and its dual (\ref{eq:nearhorizonelectromagneticpotdual1KS}). Moreover, after finding the gauge field in the near-horizon limit, we can also find the dilaton and axion field in this limit which are given as
\begin{eqnarray}
	\phi_+ =\exp\left( \frac{(m+d)^2 + (k+a\cos\theta)^2}{\varrho_+^2} \right), ~~~~
	\chi_+ = 2\frac{mk-da\cos\theta}{(m +d)^2 + (k+a\cos\theta)^2}. \label{eq:nearhorizonaxion}\
\end{eqnarray}
It is worth noting that these near-horizon forms of the fields are needed  in the calculation of the central charge. However, basically their contribution will not appear directly to the central term. 

In order to find the isometry group of the near-horizon extremal metric (\ref{eq:extremalmetricnew}), one can apply the Killing equation and solve it. In the end,  by solving Killing equation, it is found that the spacetime metric (\ref{eq:extremalmetricnew}) possesses $ SL(2,R) \times U(1) $ isometry group which are represented by the following Killing fields
\begin{equation}
	\zeta_0 = \partial_\phi , \label{eq:rotationalU1}\
\end{equation}
which denote the rotational $U(1)$ isometry and
\begin{equation}
	X_1 = \partial_t, ~~~ X_2 = t \partial_t - r \partial_r, ~~~ X_3 = \left(\frac{1}{2r^2}+\frac{t^2}{2} \right)\partial_t -t r \partial_r - \frac{e}{r}\partial_\phi .\label{eq:isometrynearhorizon}
\end{equation}
which generates $ SL(2,R)$. The Killing vectors (\ref{eq:rotationalU1}) and (\ref{eq:isometrynearhorizon}) fully denote an enhanced $ SL(2,R) \times U(1) $ isometry group. This isometry resembles the isometry of AdS$_2 $ spacetime, so the asymptotic symmetry group as used by Brown and Henneaux \cite{brown1986} can be employed to compute the central charge. Note that only $U(1)$ that can be extended to a Virasoro algebra while the $SL(2,R)$ is taken to be frozen at extremality \cite{Compere2017}. Enticingly, for non-extremal black hole, one can construct similar $ SL(2,R) \times U(1) $ isometry where $SL(2,R)$ isometry can be extended to Kac-Moody algebra \cite{AggarwalCastroJHEP2020}. In this case, one can extend the calculation to the entropy in warped CFT. However, we will not consider this case within this paper.

\subsection{Charges}

The approach of Brown and Henneaux \cite{brown1986} can be employed to find the central charge of the holographic dual CFT description of an extremal rotating black hole in EMDA theory. To compute the charges that associate with asymptotic symmetry group (ASG) of near-horizon extremal dyonic Kerr-Sen black hole, we should consider all possible contributions from all different fields in the action. Nonetheless, it has been pointed out by Comp{\'e}re \cite{Compere2017} and also shown in Ref. \cite{Ghezelbash2009} for Kerr-Sen black hole that the contributions from electromagnetic and scalar fields, except gravity are zero. Their contributions emerge only through the black hole's parameters in the central term from gravity part. So, asymptotic symmetry of the general black hole family in EMDA theory includes diffeomorphisms $ \xi $ that satisfy
\begin{eqnarray}
	\delta_\xi g_{\mu\nu} = \mathcal{L}_\xi g_{\mu\nu} =  \xi^\sigma (\partial_\sigma g_{\mu\nu}) + g_{\mu\sigma}(\partial _\nu \xi^\sigma)+ g_{\sigma \nu}(\partial _\mu \xi^\sigma), \ 
\end{eqnarray}
where the metric deviation is denoted by $ \delta_\xi g_{\mu\nu} = h_{\mu\nu} $. The associated conserved charge is
\begin{eqnarray}
	Q_{\xi} = \frac{1}{8\pi}\int_{\partial\Sigma} k^{g}_{\zeta}[h;g].
\end{eqnarray}
The given integral is over the boundary of a spatial slice. The contribution of the metric tensor on the central charge is given explicitly by
\begin{eqnarray}
	k^{g}_{\zeta}[h;g] &=& -\frac{1}{4}\epsilon_{\rho\sigma\mu\nu} \left\{ \zeta ^{\nu} D^{\mu} h - \zeta ^{\nu} D_{\lambda} h^{\mu \lambda} + \frac{h}{2} D^{\nu}\zeta^{\mu} - h^{\nu \lambda} D_{\lambda}\zeta^{\mu} + \zeta_{\lambda}D^{\nu}h^{\mu \lambda} \right. \nonumber\\
	& & \left. +\frac{h^{\lambda \nu}}{2}\left(D^\mu \zeta_\lambda + D_\lambda \zeta^\mu \right) \right\} dx^\rho \wedge dx^\sigma . \label{eq:kgrav} \
\end{eqnarray}
We should note that the last two terms in Eq. (\ref{eq:kgrav}) vanish for an exact Killing vector and an exact symmetry, respectively. The charge $ Q_{\zeta} $ generates symmetry through the Dirac brackets. The ASG possesses algebra which is given by the Dirac bracket algebra of the following charges \cite{BarnichBrandt2002}
\begin{eqnarray}
	\{ Q_{\zeta},Q_{\bar{\zeta}} \}_{DB} &=& \frac{1}{8\pi}\int k^{g}_{\zeta}\left[\mathcal{L}_{\bar{\zeta}}g;g \right] \nonumber\\
	&=& Q_{[\zeta,\bar{\zeta}]} + \frac{1}{8\pi}\int k^{g}_{\zeta}\left[\mathcal{L}_{\bar{\zeta}}\bar{g};\bar{g}\right] \label{eq:charges}.\
	\
\end{eqnarray}

In order to employ the ASG, we need to specify the boundary conditions on the metric deviations $h_{\mu\nu}$. The boundary conditions are imposed in order to produce finite and integrable charges. There is not necessarily a unique
set of consistent boundary conditions. Therefore, we adopt the boundary conditions such in most Kerr/CFT correspondence articles. In the basis $ (t,r,\theta, \phi) $, we impose the following boundary conditions
\begin{eqnarray}
	h_{\mu \nu} \sim \left(\begin{array}{cccc}
		\mathcal{O}(r^2) & \mathcal{O}\left(\frac{1}{r^2}\right) &  \mathcal{O}\left(\frac{1}{r}\right) &  \mathcal{O}(1) \\
		&  \mathcal{O}\left(\frac{1}{r^3}\right) &  \mathcal{O}\left(\frac{1}{r^2}\right) &  \mathcal{O}\left(\frac{1}{r}\right) \\
		&  & \mathcal{O}\left(\frac{1}{r}\right) &  \mathcal{O}\left(\frac{1}{r}\right)\\
		&  &  &  \mathcal{O}(1)\
	\end{array} \right).\label{eq:gdeviation}
\end{eqnarray}
The most general diffeomorphism symmetry that preserves such boundary conditions ({\ref{eq:gdeviation}) is generated by the following Killing vector field
	\begin{eqnarray}
		\zeta = \left\{c_t + \mathcal{O}\left(r^{-3}\right) \right\}\partial _t + \left\{-r\epsilon '(\phi) + \mathcal{O}(1) \right\}\partial _r + \mathcal{O}\left(r^{-1}\right)\partial _\theta  + \left\{\epsilon(\phi) + \mathcal{O}\left(r^{-2}\right) \right\}\partial _\phi ,\
	\end{eqnarray}
	where $ c_t $ is an arbitrary constant and the prime $ (') $ denotes the derivative respect to $ \phi $. This ASG contains one copy of the conformal group of the circle which is generated by
	\begin{eqnarray}
		\zeta_\epsilon = \epsilon(\phi)\partial_\phi - r\epsilon '(\phi)\partial_r ,\label{eq:killingASG}
	\end{eqnarray}
	that will be the part of the near-horizon extremal metric. We know that the azimuthal coordinate is periodic under the rotation $ \phi \sim \phi+2\pi $. Hence we may define $ \epsilon_z= -e^{-iz \phi} $ and $ \zeta_\epsilon =\zeta_\epsilon(\epsilon_z ) $. By the Lie bracket, the symmetry generator (\ref{eq:killingASG}) satisfies the Witt algebra
	\begin{eqnarray}
		i[\zeta_y, \zeta_z]_{LB} = (y-z)\zeta_{y+z}.
	\end{eqnarray}
	Then by defining
	\begin{eqnarray}
		Q_{\zeta} \equiv L_{z} - x \delta_{z,0},
	\end{eqnarray}
	on (\ref{eq:charges}) where $ x $ is a free parameter which will not change the central charge, we obtain the conserved charges algebra in quantum form, such that
	\begin{eqnarray}
		\left[L_y, L_z \right] = (y-z) L_{y+z} + \frac{c_L}{12} (y^2-1)\delta_{y+z, 0}. 
	\end{eqnarray}
	From the algebra above, we can read-off the value of the left-moving central charge for the near-horizon extremal dyonic Kerr-Sen black hole. It is obtained that
	\begin{equation}
		c_L = 12am=12J, \label{eq:cLgdyonicKS}\
	\end{equation}
This result is identical to the central charge of Kerr-Sen \cite{Ghezelbash2009} and Kerr \cite{GuicaPRD2009} black holes. However, we need to note that the relation between mass, spin, and electromagnetic charges is different with to those of Kerr-Sen and Kerr black holes. For extremal dyonic Kerr-Sen black hole, since the mass is given in Eq. (\ref{mext}). Each branch corresponds to CFT with the central charge $c_{L}=12am_{\pm}$, respectively. Recall that these branches will reduce into one only when the spin approaches zero denoting that $c_L$ also vanishes. This means that the entropy from CFT will also vanish. Vanishing $c_L$ is trivial and exactly denotes that the Virasoro algebra reduces to the classical Witt algebra. 

\subsection{Temperatures}
	Before going to match the entropies, it is required to calculate the corresponding temperatures. In order to do so, we need to employ the analog of the Hartle-Hawking vacuum, i.e. Frolov-Thorne vacuum that has been used in the Kerr/CFT correspondence \cite{GuicaPRD2009} because the angular momentum and other thermodynamic quantities are included within this vacuum. Now, we may apply the first law of black hole thermodynamics for rotating dyonic Kerr-Sen black holes (\ref{eq:thermoKS}) where the extremal condition satisfies
	\begin{eqnarray}
		dM= \Omega_H^{ex} dJ +\Phi_H^{ex} dQ +\Psi_H^{ex} dP , \label{eq:thermoKSAdSex}
	\end{eqnarray}
	since $T^{ex}_H=0$. So, we can write
	\begin{equation}
		T_H dS = -\left[ (\Omega_H - \Omega_H^{ex})dJ + (\Phi - \Phi^{ex}) dQ + (\Psi - \Psi^{ex})dP \right]. \label{eq:constrainofthermo}
	\end{equation}
	For such constrained variations (\ref{eq:constrainofthermo}), we may construct
	\begin{eqnarray}
		dS = \frac{dJ}{T_L}+\frac{dQ}{T_q}+\frac{dP}{T_p} .\
	\end{eqnarray}
	
	For Kerr black hole, it is considered a quantum scalar field with eigenmodes of the asymptotic energy $ E $ and angular momentum $ J $ which are given by the following form
	\begin{eqnarray}
		\tilde{\Phi} = \sum_{E,J,s} \tilde{\phi} _{E,J,s} e^{-i E \hat{t} + i J \hat{\phi}} f_s(\hat{r},\theta),
	\end{eqnarray}
	In order to transform this to near-horizon quantities and take the extremal limit, we note that in the near-horizon coordinates (\ref{eq:extremaltransformation}) we have
	\begin{eqnarray}
		e^{-i E \hat{t} + i J \hat{\phi}} = e^{-i (E-\Omega_H^{ex} J )t r_0/\epsilon + i J \phi} = e^{-in_R t + in_L \phi},
	\end{eqnarray}
	where
	\begin{eqnarray}
		n_R = (E-\Omega_H^{ex} J)r_0/\epsilon, ~~~n_L = J. \label{eq:nrnl}
	\end{eqnarray}
	But this is only suitable when there is no contribution of other thermodynamical potentials. Using the fact that any system possesses density of state $\rho=e^S$, where $S$ is the entropy and the fact that there are thermodynamic potentials coming from electromagnetic fields, we may extend Eq. (\ref{eq:nrnl}) to 
	\begin{eqnarray}
		n_R = (E-\Omega_H^{ex} J- \Phi^{ex} Q- \Psi^{ex} P ) r_0/\epsilon, ~~~n_L = J. \label{eq:nRnLnew}
	\end{eqnarray}
	The density matrix in the asymptotic energy, angular momentum, electric charge, magnetic charge, and pressure eigenbasis now has the Boltzmann weighting factor
	\begin{eqnarray}
		e^{-\left( \frac{E - \Omega_H J - \Phi^{ex} Q- \Psi^{ex} P }{T_H}\right) } =e^{-\frac{n_R}{T_R}-\frac{n_L}{T_L}-\frac{Q}{T_q}-\frac{P}{T_p} }. \label{eq:Boltzmannweightingfactor}\
	\end{eqnarray}
	
	When the trace over the modes inside the horizon is taken, the Boltzmann weighting factor will be a diagonal matrix. We can compare the Eqs. (\ref{eq:nRnLnew}) and (\ref{eq:Boltzmannweightingfactor}) to obtain the definition of the CFT temperatures, such that
	\begin{eqnarray}
		&&T_R = \frac{T_H r_0}{\epsilon}\bigg|_{ex} , ~~~ T_L = - \frac{\partial T_H/\partial r_+}{\partial \Omega_H / \partial r_+}\bigg|_{ex},~~~T_q = - \frac{\partial T_H/\partial r_+}{\partial \Phi / \partial r_+}\bigg|_{ex}, ~~~ T_p = - \frac{\partial T_H/\partial r_+}{\partial \Psi / \partial r_+}\bigg|_{ex} . \label{eq:generalCFTtemperature}\
	\end{eqnarray}
	However, we only need $T_L$ and $T_R$ in order to compute the Cardy entropy. It is obviously seen that $T_R=0$ for extremal black holes. We also have
	\begin{equation}
		T_L =  \frac{m^2-d^2-k^2+a^2}{4\pi a m}. \label{eq:TLKS}\
	\end{equation}
We have obtained the left-moving temperatures for dyonic Kerr-Sen. Moreover, as mentioned before that this extremal black hole corresponds with two different branches of central charge, $12am_\pm$, the left-moving temperature also behaves the same $T_L =  \frac{m_\pm^2-d^2-k^2+a^2}{4\pi a m_\pm}$. From this temperature, we can also obtain the left-moving temperature of Kerr-Sen black hole by setting $ k=0 $. When we turn off both electromagnetic charges, the temperature for Kerr black hole is recovered \cite{GuicaPRD2009}.

\subsection{Entropy Matching}
	
We have discussed the existence of an asymptotic Virasoro algebra at the boundary in infinity  of the near-horizon extremal geometry. By following semi-classical quantization rules, the operators that define quantum gravity with the given boundary conditions form a Virasoro algebra. We have also provided that scalar quantum fields in the analogue of the Frolov-Thorne vacuum restricted to extremal excitations having the non-vanishing left-moving temperature. Since we identify the left-sector with excitations along $\partial_\phi$ and the $SL(2,R)$ isometry as the right sector is frozen, the states are described by a thermal density matrix with temperatures $ T_L, T_q$, and $ T_p $. As mentioned in \cite{Compere2017}, $T_q, T_p $ are better interpreted as the CFT chemical potentials as $\mu_L^{q}=-T_L/T_q $ and $ \mu_L^{p}=-T_L/T_p$. 

Given all quantities that we need in Cardy formula, we can now compute it in (\ref{eq:Cardyentropy}). It is remarkable that, surprisingly, using the left-moving central charge (\ref{eq:cLgdyonicKS}) and temperature (\ref{eq:TLKS}) reproduces the Bekenstein-Hawking entropy (\ref{eq:SBHKS}) for extremal dyonic Kerr-Sen black hole
	\begin{equation}
		S_{CFT}=\pi(m^2 -d^2 -k^2 +a^2)=S_{BH}. \label{eq:SCFTKSAdSshift}
	\end{equation}
Note that for given values of $a, p, q$, there exist two branches of extremal black hole's entropy with mass $m_\pm$ and central charges $12am_{\pm}$. Simultaneously, we have also reproduced the Bekenstein-Hawking entropy for Kerr-Sen black hole by turning off $p$. Again, when we assume that $q=p=0$, it reduces to the entropy of extremal Kerr black hole. For $a\rightarrow 0$ but keeping $p,q\neq 0$, this Cardy entropy will vanish showing the entropy of extremal dyonic GMGHS black hole. This matching completes the conjecture of Kerr/CFT correspondence for extremal dyonic Kerr-Sen black holes in EMDA theory. This is clearly not a coincidence and shows that the dyonic Kerr-Sen black hole is dual to 2D CFT represented by non-vanishing left-moving sector. 

\section{Kerr/CFT for Gauged Dyonic Kerr-Sen Black Hole}
\label{sec:KerrCFTgauge}
\subsection{Spacetime metric and thermodynamics}
For the gauged case, the corresponding Lagrangian density of the EMDA theory (\ref{eq:Lagrangian}) is given by \cite{WuWuWuYuPRD2021,ChowCompereCQG2014,ChowComperePRD2014}
	\begin{eqnarray}
		\mathcal{L}_{gauged} =\mathcal{L} + \sqrt{-g}\frac{4+e^{-\phi}+e^{\phi}(1+\chi^2)}{l^2}.\label{eq:Lagrangian1}
	\end{eqnarray}
The gauged version of the dyonic Kerr-Sen black hole or the Kerr-Sen-AdS black is also presented in a simple form in Ref.~\cite{WuWuWuYuPRD2021} which is also the solution of above Lagrangian density. In the shifted radial coordinate, the spacetime metric including the electromagnetic potential, its dual, dilaton field, and axion field, are given by
\begin{eqnarray}
	ds^2 & = & - \frac{\Delta}{\varrho^2 } \hat {X}^2+  \frac{\varrho ^2}{\Delta}d\hat{r}^2 + \frac{\varrho ^2}{\Delta_\theta} d\theta^2 + \frac{\Delta_\theta \sin^2\theta}{\varrho ^2}\hat{Y}^2,\label{eq:dKSAdSmetricshift} \\
	\textbf{A} &=& \frac{q(\hat{r}+d-p^2/m)}{\varrho^2}\hat{X}-\frac{p\cos\theta}{\varrho^2}\hat{Y}, ~~~	\textbf{B} = \frac{p(\hat{r}+d-p^2/m)}{\varrho^2}\hat{X}+\frac{q\cos\theta}{\varrho^2}\hat{Y}, \label{eq:BdKSAdSshift}\\
	e^{\phi} &=& \frac{(\hat{r}+d)^2+(k+a\cos\theta)^2}{\varrho^2}, ~~~
	\chi = 2\frac{k\hat{r}-da\cos\theta}{(\hat{r}+d)^2+(k+a\cos\theta)^2}, \label{eq:axiondKSAdSshift}
\end{eqnarray}
respectively, where
\begin{eqnarray}
	\hat{X} &=& d\hat{t} - a \sin^2\theta \frac{d\hat{\phi}}{\Xi}, ~~~ \hat{Y} = ad\hat{t}- (\hat{r}^2-d^2-k^2+a^2 ) \frac{d\hat{\phi}}{\Xi}, \nonumber\\
	\Delta &=& (\hat{r}^2-d^2-k^2+a^2)\left(1+\frac{\hat{r}^2-d^2 -k^2}{l^2} \right)-2m\hat{r}+p^2 +q^2, \nonumber\ \label{DelAdSshifted}\\
	\Delta_\theta &=& 1-\frac{a^2}{l^2}\cos^2\theta, ~~\Xi =1-\frac{a^2}{l^2}, ~~~ \varrho^2 =\hat{r}^2-d^2 -k^2 + a^2\cos^2\theta.\
\end{eqnarray}

The dyonic Kerr-Sen-AdS black hole satisfies the following thermodynamic relation
\begin{eqnarray}
	dM= T_H dS+\Omega_H dJ +\Phi_H dQ +\Psi_H dP + Vd\mathcal{P}, \label{eq:thermoKSAdS}
\end{eqnarray}
When cosmological constant or the gauge coupling constant $l$ arises as the vacuum expectation value, we can include this parameter in the first law of thermodynamics for black holes \cite{CveticPopePRD2011}. As common rotating AdS black holes, the cosmological constant can be considered as the source of the pressure on the black holes. Hence, it gives rise to another thermodynamic quantity. The quantities on above equations are given by
\begin{equation}
	M= \frac{m}{\Xi}, ~~~ J = \frac{ma}{\Xi}, ~~~ Q= \frac{q}{\Xi}, ~~~ P= \frac{p}{\Xi},~~~	T_H = \frac{r_+(2r_+^2-2d^2 -2k^2+a^2 +l^2)-ml^2}{2\pi (r_+^2 -d^2-k^2 +a^2)l^2}, \label{eq:THKSAdSshift}
\end{equation}
\begin{equation}
	S_{BH}=\frac{\pi}{\Xi}(r_+^2 -d^2 -k^2 +a^2), ~~~
	\Omega_H = \frac{a\Xi}{r_+^2 -d^2 -k^2 +a^2}, ~~~
	\Phi_H = \frac{q(r_+ +d -p^2/m)}{r_+^2 -d^2 -k^2 +a^2}, \label{eq:potKSAdSshift}
\end{equation}
\begin{equation}
\Psi_H = \frac{p(r_+ +d -p^2/m)}{r_+^2 -d^2 -k^2 +a^2}, ~~~	V= \frac{4}{3}r_+ S, ~~~\mathcal{P}=\frac{3}{8\pi l^2}. \label{eq:VdKSAdSshift}
\end{equation}
where those are mass, angular momentum, electric charge, magnetic charge, Hawking temperature, Bekenstein-Hawking entropy, angular velocity, electric potential, magnetic potential, volume and pressure. 

Since there exists the cosmological constant, this black hole possesses more than two horizons as we can see from $\Delta$ which is a quartic function. This means that there also exists cosmological horizons. From this dyonic Kerr-Sen-AdS black hole solution, we can find several solutions by taking certain limits. In order to find Kerr-Sen-AdS black hole, one can turn off $p=0$ that will cause $k=0$. This implies $\Psi_H=0$. When one consider equal charges $q=p$, this will result in $d=0$ or no dilaton charge. On the other hand, we can find that $\Phi_H =\Psi_H$. Another fascinating property of this solution is we can find the superentropic solution of dyonic Kerr-Sen-AdS black hole by taking $a\rightarrow l$. Nevertheless, whether it is superentropic or not is defined by the value of $l^2$ over $(d^2 +k^2)$.

\textcolor{black}{Next, we consider the extremal limit for the gauged counterparts of the dyonic Kerr-Sen black hole with $\Delta'=0$. We find
\begin{equation}
	m= \frac{r_+}{l^2}(2r_+^2-2d^2-2k^2+a^2+l^2).  \label{meqn}
\end{equation}
As a result, the horizon is given by
\begin{eqnarray}
	r_+^2 = \frac{2d^2 +2k^2-a^2-l^2 + \sqrt{x_1^4 -16x_2^4+2x_3^4}}{6},\label{eq:EHdKSAdS}\
\end{eqnarray}
where $x_1^4=a^4 +16d^4+16k^4+l^4$, $x_2^4=a^2 d^2+a^2 k^2+d^2 l^2+k^2l^2-2d^2k^2$, and $x_3^4 = 7a^2l^2 + 6l^2q^2 + 6l^2p^2$. All the horizons now coincide into one. Similar to the ungauged case, the Hawking temperature vanishes while other thermodynamic quantities remain non-zero. Hence, there are possibly more than two branches of extremal black holes corresponding to roots of (\ref{meqn}) since $d,k \sim 1/m$. The physical conditions for the existence of extremal black holes are $x_1^4 -16x_2^4+2x_3^4 \geq 0, r_{+}^{2}>0$ and $m>0$.
}

\subsection{Dual CFT of Dyonic Kerr-Sen-AdS Black hole}
The near-horizon form of dyonic Kerr-Sen-AdS black hole can be obtained using similar transformation (\ref{eq:extremaltransformation}). In order to study the near-horizon extremal region, we need to approximate $\Delta$ in terms of event horizon $r_+$. In the near-horizon of extremal black holes, the function $ \Delta $ takes form \cite{SaktiEPJPlus2019}
\begin{equation}
	\Delta = (\hat{r}-r_+)^2 \upsilon + \mathcal{O}\left((\hat{r}-r_+)^3 \right),\
\end{equation}
where the function $\upsilon$ is given by
\begin{equation}
	\upsilon = \frac{\Delta''(r_+)}{2}= 1 + \frac{6r_+^2-2d^2-2k^2+a^2}{l^2}.\label{eq:upsilon}
\end{equation}
We can obtain the near-horizon extremal metric of Eq. (\ref{eq:dKSAdSmetricshift}). By additional scaling $ dt \rightarrow  dt/\upsilon $, the near-horizon extremal metric is then given by
\begin{eqnarray}
	ds^2 = \Gamma(\theta)\left(-r^2 dt^2 + \frac{dr^2}{r^2} + \alpha(\theta) d\theta ^2 \right)  +\gamma(\theta) \left(d\phi +e r dt\right)^2, \label{eq:extremalmetricnewdKSAdS}\
\end{eqnarray}
where the metric functions are given by
\begin{equation}
	\Gamma(\theta)= \frac{\varrho_+ ^2}{\upsilon} , ~~~\alpha(\theta) = \frac{\upsilon}{\Delta_\theta}, ~~~\gamma(\theta)=\frac{r_0^4 \Delta_\theta \sin^2\theta}{\varrho_+^2 \Xi^2}, ~~~
	\varrho_+^2 = r_+^2 -d^2 -k^2 + a^2\cos^2\theta, ~~~
	e = \frac{2ar_+\Xi}{r_0^2 \upsilon}. \label{eq:constant_e}
\end{equation}

In the near-horizon limit, we find that the gauge field and its dual are given by
\begin{eqnarray}
&& \textbf{A}_+ = f_a(\theta) \left(d\phi +\hat{e} r dt \right)+\frac{q\left[(r_+ +d)^2+k^2-a^2-2p^2r_+/m \right]}{\hat{e}(r_+^2-d^2-k^2+a^2)}d\phi, \label{eq:nearhorizonelectromagneticpot1}\\
&& \textbf{B}_+ = f_b(\theta) \left(d\phi + \hat{e} r dt \right)+\frac{p\left[(r_+ +d)^2+k^2-a^2-2p^2r_+/m \right]}{\hat{e}(r_+^2-d^2-k^2+a^2)}d\phi,  ~~~ \hat{e} = \frac{2ar_+\Xi}{r_0^2},
	\label{eq:nearhorizonelectromagneticpotdual1}\
\end{eqnarray}
where the functions $f(\theta)$ for each field are given by
\begin{eqnarray}
	f_a(\theta) = \frac{q\left[\frac{2r_+p^2}{m}-(r_+ + d)^2-k^2+a^2\cos^2\theta\right] +2apr_+\cos\theta}{2ar_+\Xi \varrho_+^2}r_0^2, \label{eq:nearhorizonelectromagneticpot2}\\
	f_b(\theta) = \frac{p\left[\frac{2r_+p^2}{m}-(r_+ + d)^2-k^2+a^2\cos^2\theta\right] -2aqr_+\cos\theta}{2ar_+\Xi \varrho_+^2}r_0^2,\
	\label{eq:nearhorizonelectromagneticpotdual2}
\end{eqnarray}
where we can gauge away the second term on the electromagnetic potential (\ref{eq:nearhorizonelectromagneticpot1}) and its dual (\ref{eq:nearhorizonelectromagneticpotdual1}). Moreover, after finding the gauge field in the near-horizon limit, we can also find the dilaton and axion field in this limit which are given as
\begin{eqnarray}
	e^{\phi_+} = \frac{(r_+ +d)^2 + (k+a\cos\theta)^2}{\varrho_+^2}, ~~~\chi_+ = 2\frac{kr_+ -da\cos\theta}{(r_+ +d)^2 + (k+a\cos\theta)^2}. \label{eq:nearhorizonaxion}\
\end{eqnarray}

In order to compute the central charge of dyonic Kerr-Sen-AdS black hole, we can also employ the similar ASG calculation to the near-horizon extremal metric (\ref{eq:extremalmetricnewdKSAdS}) since we obtain the identical metric form and isometry. From the same lengthy calculation, we can obtain the left-moving central charge for the near-horizon extremal dyonic Kerr-Sen-AdS black holes from EMDA theory. It is obtained that
\begin{equation}
	c_L = \frac{12ar_+}{\upsilon}, \label{eq:cLgeneral}\
\end{equation}
where the event horizon $\upsilon$ and $r_+$ are given in (\ref{eq:upsilon}) and (\ref{eq:EHdKSAdS}), respectively.
We can also recover the central charge of Kerr-Sen-AdS and Kerr-Sen-AdS black holes with vanishing axion charge by setting $p=0$ and $p=q$, respectively. It is also clear that by turning off $p,q$, we can find the central charge of Kerr-AdS black hole \cite{Lu2009}. 

Similarly, before going to match the entropy, it is required to calculate the corresponding temperatures. For gauged case, the presence of the cosmological constant can be considered as the additional thermodynamic quantity in the thermodynamic equation. Hence, we can write the first law of black hole thermodynamics for this black hole where the extremal condition satisfies
\begin{eqnarray}
	dM= \Omega_H^{ex} dJ +\Phi_H^{ex} dQ +\Psi_H^{ex} dP + V^{ex}d\mathcal{P}, \label{eq:thermoKSAdSex}
\end{eqnarray}
since $T^{ex}_H=0$. We are required to write that
\begin{equation}
	T_H dS = -\left[ (\Omega_H - \Omega_H^{ex})dJ + (\Phi - \Phi^{ex}) dQ + (\Psi - \Psi^{ex})dP + (V - V^{ex})d\mathcal{P}\right]. \label{eq:constrainofthermoAdS}
\end{equation}
For such constrained variations (\ref{eq:constrainofthermoAdS}), we may construct
\begin{eqnarray}
	dS = \frac{dJ}{T_L}+\frac{dQ}{T_q}+\frac{dP}{T_p}+\frac{d\mathcal{P}}{T_\mathcal{P}} .\
\end{eqnarray}
The last term in above equation is related to the cosmological constant.

Similar to the dyonic Kerr-Sen black hole, there exist temperatures conjugate to electric and magnetic charges aside from the right- and left-moving temperatures. The existence of cosmological constant is related to the presence of another temperature conjugate to the cosmological pressure which is defined as $ T_{\mathcal{P}}= - \frac{\partial T_H/\partial r_+}{\partial V / \partial r_+}\bigg|_{ex}$,  although we only need the the left-moving temperature in this case. The explicit expression of the left-moving temperature is
\begin{equation}
	T_L =  \frac{\upsilon(r_+^2-d^2-k^2+a^2)}{4\pi a r_+ \Xi}. \label{eq:TLgeneral}\
\end{equation}
From this temperature, we can recover the left-moving temperatures for Kerr-Sen-AdS and Kerr-Sen-AdS with vanishing axion charge similarly with the central charge. When we turn off both electromagnetic charges, the temperatures for Kerr-AdS black hole is then recovered.

We have calculated the quantities that we need in Cardy formula, we can now compute it in (\ref{eq:Cardyentropy}). It is remarkably found that for extremal dyonic Kerr-Sen-AdS black hole possesses the following CFT entropy
\begin{equation}
	S_{CFT}=\frac{\pi}{\Xi}(r_+^2 -d^2 -k^2 +a^2). \label{eq:SCFTKSAdSshift}
\end{equation}
This result matches with the Bekenstein-Hawking entropy for dyonic Kerr-Sen-AdS black hole. The entropy of Kerr-Sen-AdS black hole and Kerr-Sen-AdS black hole with vanishing axion charge can also be found from this. When we assume that $q=p=0$, it reduces to the entropy of extremal Kerr-AdS black hole \cite{Lu2009}. It is worth noting that for Kerr-Newman-AdS black holes, there exists second dual CFT \cite{Hartman2009} in which can be studied when considering $a\rightarrow 0$ and then proposing the electromagnetic field as the part of the geometry. In this second dual CFT, the temperatures $T_q$ and $T_p$ are useful to reveal the dual CFT of Reissner-Nordstr{\"o}m black hole family. Nonetheless, for black hole solutions in EMDA theory, in this case is the Kerr-Sen black hole \cite{GhezelbashSiahaanCQG2013}, the second dual CFT fails to be shown. Hence, we will not further consider the second dual CFT for the dyonic Kerr-Sen-AdS black hole family in this paper.

\section{Ultraspinning dyonic Kerr-Sen-AdS black hole}
\label{sec:KerrCFTultra}
\subsection{Spacetime metric and thermodynamics}
Another interesting spacetime we want to study is the ultraspinning dyonic Kerr-Sen-AdS black hole. In this circumstance, we consider ultraspinning limit $a\rightarrow l$. This metric has been given in Ref.~\cite{WuWuWuYuPRD2021} while herein we consider the spacetime metric in shifted radial coordinate. In order to find the ultraspinning version of the dyonic Kerr-Sen-AdS black hole, we need to re-define the coordinate $\hat{\phi}$ as  $\hat{\phi} \rightarrow \hat{\phi}\Xi $ to exclude the conical singularity after taking ultraspinning limit. Since the azimuthal coordinate becomes non-compact, we need to compactify it as $\hat{\phi} \rightarrow \hat{\phi} + \mu$ where $\mu$ is not 2$\pi$ and dimensionless. As explained in Ref.~\cite{HennigarMannPRL2015}, $\mu$ is proposed as another thermodynamic quantity or the chemical potential. The resulting metric, electromagnetic potential, dual of the electromagnetic potential, dilaton and axion fields are given by
\begin{eqnarray}
	ds^2 & = & - \frac{\Delta}{\varrho^2 } \hat {X}^2+  \frac{\varrho ^2}{\Delta}d\hat{r}^2 + \frac{\varrho ^2}{\sin^2\theta} d\theta^2 + \frac{\sin^4\theta}{\varrho ^2}\hat{Y}^2,\label{eq:dKSAdSSmetric} \\
	\textbf{A} &=&  \frac{q(r+d-p^2/m)}{\varrho^2}\hat{X}-\frac{p\cos\theta}{\varrho^2}\hat{Y}, ~~~	\textbf{B} = \frac{p(r+d-p^2/m)}{\varrho^2}\hat{X}+\frac{q\cos\theta}{\varrho^2}\hat{Y}, \label{eq:BdKSAdSS}\\
	e^{\phi} &=& \frac{(\hat{r}+d)^2+(k+l\cos^2\theta)}{\varrho^2},~~~
	\chi = 2\frac{k\hat{r}-dl\cos\theta}{(\hat{r}+d)^2+(k+l\cos^2\theta)}, \label{eq:axiondKSAdSS}
\end{eqnarray}
respectively, where
\begin{equation}
	\hat{X} = d\hat{t} - l \sin^2\theta d\hat{\phi}, ~~ \hat{Y} = ld\hat{t}- (\hat{r}^2-d^2-k^2+l^2 ) d\hat{\phi}.\nonumber\
\end{equation}
\begin{equation}
	\Delta = \frac{(\hat{r}^2-d^2-k^2+l^2)^2}{l^2} -2m\hat{r}+p^2 +q^2, ~~~
	\varrho^2 =\hat{r}^2-d^2-k^2+l^2\cos ^2\theta.\nonumber\
\end{equation}
For the ultraspinning black hole, the thermodynamic quantities satisfy the following relation
\begin{eqnarray}
	dM= T_H dS+\Omega_H dJ +\Phi_H dQ +\Psi_H dP + Vd\mathcal{P} + K d\mu, \label{eq:thermoKSAdSS}
\end{eqnarray}
where $K$ is the conjugate of $\mu$ or the conjugate of chemical potential. The thermodynamic quantities of this black hole are given by
\begin{equation}
	M= \frac{m \mu}{2\pi}, ~~~ J = \frac{ml\mu}{2\pi}, ~~~ Q= \frac{q \mu}{2\pi}, ~~~ P= \frac{p\mu}{2\pi}, ~~~
	T_H = \frac{2r_+(r_+^2-d^2 -k^2+l^2)-ml^2}{2\pi (r_+^2 -d^2-k^2 +l^2)l^2}, \label{eq:THKSAdSS}
\end{equation}
\begin{equation}
	S_{BH}=\frac{\mu}{2}(r_+^2 -d^2 -k^2 +l^2), ~~~
	\Omega_H = \frac{l}{r_+^2 -d^2 -k^2 +l^2}, 	~~~ \Phi_H = \frac{q(r_+ +d -p^2/m)}{r_+^2 -d^2 -k^2 +l^2}, \label{eq:angvelKsAdSSshift}
\end{equation}
\begin{equation}
\Psi_H = \frac{p(r_+ +d -p^2/m)}{r_+^2 -d^2 -k^2 +l^2}, ~~~
	V= \frac{2\mu}{3}r_+ (r_+^2 -d^2 -k^2 +l^2), ~~~
	K=m \frac{l^2 -(r_+^2-d^2-k^2)}{4\pi(r_+^2 -d^2 -k^2 +l^2)}. \label{eq:VdKSAdSSshift}
\end{equation}

\textcolor{black}{Again, the extremal limit can be found by using conditions $\Delta =0$ and $\Delta'=0$,
\begin{equation}
	m= \frac{2r_+}{l^2}(r_+^2-d^2-k^2+l^2).  \label{meqn1}
\end{equation}
The event horizon of this extremal ultraspinning version is located at
\begin{eqnarray}
	r_+^2 =\frac{d^2 +k^2-l^2+\sqrt{4l^4+4d^4+4k^4-8l^2d^2-8l^2k^2 +8d^2k^2+3l^2q^2+3l^2p^2}}{3}.\label{eq:EHdKSAdSS}
\end{eqnarray}
In this extremal ultraspinning limit, the Hawking temperature also vanishes. There are possibly more than two branches of extremal ultraspinning black holes corresponding to roots of (\ref{meqn1}). The physical conditions for the existence of extremal black holes are $4l^4+4d^4+4k^4-8l^2d^2-8l^2k^2 +8d^2k^2+3l^2q^2+3l^2p^2 \geq 0, r_{+}^{2}>0$ and $m>0$.
}

For the ultraspinning version of the dyonic Kerr-Sen-AdS black hole, the near-horizon coordinate transformations are identical. The near-horizon extremal geometry of ultraspinning dyonic Kerr-Sen-AdS black hole is given by Eq. (\ref{eq:extremalmetricnewdKSAdS}) with the following functions
\begin{equation}
	\Gamma(\theta)= \frac{\varrho_+ ^2}{\upsilon} , ~~~\alpha(\theta) = \frac{\upsilon}{\sin^2\theta}, ~~~\gamma(\theta)=\frac{r_0^4\sin^4\theta}{\varrho_+^2}, ~~~
	\varrho_+^2 = r_+^2 -d^2 -k^2 + l^2\cos^2\theta, \nonumber\
\end{equation}
\begin{equation}
	\upsilon = 2 + \frac{6r_+^2-2d^2-2k^2}{l^2}, ~~~ e = \frac{2lr_+}{r_0^2 \upsilon}. \label{eq:constant_e4}
\end{equation}
The functions $f(\theta)$ for the gauge field and its dual, dilaton and axion fields in near-horizon region are given by
\begin{eqnarray}
	f_a(\theta) &=& \frac{q\left[\frac{2r_+p^2}{m}-(r_+ + d)^2-k^2+l^2\cos^2\theta\right] +2lpr_+\cos\theta}{2lr_+ \varrho_+^2}r_0^2, \label{eq:nearhorizonelectromagneticpot5}\\
	f_b(\theta) &=& \frac{p\left[\frac{2r_+p^2}{m}-(r_+ + d)^2-k^2+l^2\cos^2\theta\right] -2lqr_+\cos\theta}{2lr_+ \varrho_+^2}r_0^2,\\
	\label{eq:nearhorizonelectromagneticpotdual5}
e^{\phi_+} &=& \frac{(r_+ +d)^2 + (k+l\cos\theta)^2}{\varrho_+^2},~~~
	\chi_+ = 2\frac{kr_+-dl\cos\theta}{(r_+ +d)^2 + (k+l\cos\theta)^2}. \label{eq:nearhorizonaxion4}\
\end{eqnarray}
For this ultraspinning version, the near-horizon geometry of the black hole with metric potential (\ref{eq:constant_e4}) also possesses $SL(2,R)\times U(1)$ isometry group with similar Killing vector fields (\ref{eq:isometrynearhorizon}) and (\ref{eq:rotationalU1}).

\subsection{Dual CFT of Ultraspinning Dyonic Kerr-Sen-AdS Black hole}
By employing ASG calculation, for ultraspinning dyonic Kerr-Sen-AdS, the central charge is given by
\begin{equation}
	c_L= \frac{6r_+\mu l}{\pi\upsilon}, \label{eq:cLultra}
\end{equation}
which interestingly depends on chemical potential $\mu$, where $\upsilon $ and $r_+$ are given  in Eqs. (\ref{eq:constant_e4}) and (\ref{eq:EHdKSAdSS}).  Even in this ultraspinning case, there are possibly more than two branches of extremal black holes related to the mass which also correspond to CFTs with different central charges. For ultraspinning black holes, since the thermodynamic relation is distinct from the normally spinning black hole because there is a chemical potential, it is required to employ the relation (\ref{eq:thermoKSAdSS}). Using the fact $T_H^{ex}=0$ in (\ref{eq:thermoKSAdSS}) for extremal case, we can construct
	\begin{eqnarray}
		T_H dS &=& -\bigg[ (\Omega_H - \Omega_H^{ex})dJ + (\Phi - \Phi^{ex}) dQ + (\Psi - \Psi^{ex})dP + (V - V^{ex})d\mathcal{P} + (K - K^{ex})d\mu \bigg], \\
		dS &=& \frac{dJ}{T_L}+\frac{dQ}{T_q}+\frac{dP}{T_p}+\frac{d\mathcal{P}}{T_\mathcal{P}}+\frac{d\mu}{T_\mu} \label{eq:constrainofthermoultra}.\
	\end{eqnarray}
	The further analysis is identical with those when we do not consider the ultraspinning case. So, we can also define another temperature conjugate to the chemical potential, $
		T_\mu = - \frac{\partial T_H/\partial r_+}{\partial K / \partial r_+}\bigg|_{ex}. $
	Nevertheless, the most significant temperature to compute Cardy formula is $T_L$ as given by
	\begin{equation}
		T_L =  \frac{\upsilon(r_+^2-d^2-k^2+l^2)}{4\pi l r_+ }. \label{eq:TLultra}\
	\end{equation}
It is remarkable that one can also find the similar matching for ultraspinning version of dyonic Kerr-Sen-AdS black hole, 
	\begin{equation}
		S_{CFT}=\frac{\mu}{2}(r_+^2 -d^2 -k^2 +l^2)=S_{BH}. \label{eq:SCFTKSAdSSshift}\
	\end{equation}
This matching so far completes the conjecture of Kerr/CFT correspondence for extremal dyonic black holes, especially in EMDA theory. This is clearly not a coincidence and shows that the dyonic Kerr-Sen black hole and its family are dual to 2D CFT represented by non-vanishing left-moving sector.
	
\section{Dyonic Kerr-Sen-AdS and Kerr-Newman-AdS in Comparison}  \label{DKSKNCom}

Here we will compare the properties from CFT of the black holes in Einstein-Maxwell theory, i.e. dyonic Kerr-Newman-AdS solution and black hole in EMDA theory, i.e. dyonic Kerr-Sen-AdS black hole. The study of superentropic black hole from rotating black holes with non-vanishing cosmological constant has been proposed in Ref.~\cite{HennigarMannPRL2015} using Kerr-Newman-AdS black hole solution. Furthermore, it has been proposed the same thermodynamical properties for similar black hole solution from the Kerr/CFT duality point of view. Using the Kerr/CFT method, authors in Ref.~\cite{Sinamuli2016} calculated the entropy of extremal ultraspinning Kerr-Newman-AdS black holes. Both papers actually do not consider the existence of dyonic solution or non-zero magnetic charge. However, their results will not be so different with dyonic solution if we just change $q^2 \rightarrow q^2 +p^2 $ where $p$ is the magnetic charge. In this section, we will compare the thermodynamical properties of the dyonic Kerr-Newman-AdS and dyonic Kerr-Sen-AdS black holes including its ultraspinning versions. Later on, we also compare with the result when we consider asymptotically dS spacetime.
	
	\subsection{Kerr-Newman-AdS black hole and its ultraspinning version}
	The dyonic Kerr-Newman-AdS black hole spacetime is given by
	\begin{eqnarray}
		ds^2 & = & - \frac{\Delta}{\varrho^2 } \hat {X}^2+  \frac{\varrho ^2}{\Delta}d\hat{r}^2 + \frac{\varrho ^2}{\Delta_\theta} d\theta^2 + \frac{\Delta_\theta \sin^2\theta}{\varrho ^2}\hat{Y}^2,\label{eq:dKNAdSmetricshift} \\
		\textbf{A} &=& \frac{-q\hat{r}}{\varrho^2}\hat{X}-\frac{p\cos\theta}{\varrho^2}\hat{Y}, ~~~	\textbf{B} = \frac{-p \hat{r}}{\varrho^2}\hat{X}+\frac{q\cos\theta}{\varrho^2}\hat{Y}, \label{eq:BdKSAdSshift}\
	\end{eqnarray}
	where $\textbf{A}$ and $\textbf{B}$ are the electromagnetic potential and its dual, respectively. Here, we have
	\begin{eqnarray}
		\hat{X} &=& d\hat{t} - a \sin^2\theta \frac{d\hat{\phi}}{\Xi}, ~~~ \hat{Y} = ad\hat{t}- (\hat{r}^2+a^2 ) \frac{d\hat{\phi}}{\Xi} \nonumber\\
		\Delta &=& (\hat{r}^2+a^2)\left(1+\frac{\hat{r}^2}{l^2} \right)-2m\hat{r}+p^2 +q^2, \nonumber\ \label{DelNAdSshifted}\\
		\Delta_\theta &=& 1-\frac{a^2}{l^2}\cos^2\theta, ~~\Xi =1-\frac{a^2}{l^2}, ~~~ \varrho^2 =\hat{r}^2 + a^2\cos^2\theta.\
	\end{eqnarray}
	
	The dyonic Kerr-Newman-AdS black hole satisfies the identical thermodynamic relation (\ref{eq:thermoKSAdSS}) as the dyonic Kerr-Sen-AdS black hole. The thermodynamic quantities for
	above metric are given by
	\begin{equation}
		M= \frac{m}{\Xi}, ~~~ J = \frac{ma}{\Xi}, ~~~ Q= \frac{q}{\Xi}, ~~~ P= \frac{p}{\Xi},~~~
		T_H = \frac{r_+(2r_+^2+a^2 +l^2)-ml^2}{2\pi (r_+^2+a^2)l^2}, \label{eq:THKNAdS}
	\end{equation}
	\begin{equation}
		S_{BH}=\frac{\pi}{\Xi}(r_+^2+a^2), ~~~
		\Omega_H = \frac{a\Xi}{r_+^2 +a^2}, ~~~
		\Phi_H = \frac{qr_+}{r_+^2 +a^2},~~~  \Psi_H = \frac{pr_+ }{r_+^2+a^2}, \label{eq:potKNAdS}
	\end{equation}
	\begin{equation}
		V= \frac{2\pi}{3\Xi}\frac{(r_+^2 +a^2)(2r_+^2l^2+a^2l^2-r_+^2a^2)+a^2l^2(q^2+p^2)}{r_+l^2\Xi}. \label{eq:VdKNAdS}
	\end{equation}
	where those are mass, angular momentum, electric charge, magnetic charge, Hawking temperature, Bekenstein-Hawking entropy, angular velocity, electric potential, magnetic potential, volume and pressure.
	The event horizon of dyonic Kerr-Newman black hole is located at
	\begin{eqnarray}
		r_+^2 = \frac{-a^2-l^2+\sqrt{a^4+l^4+14a^2l^2 + 12l^2q^2 + 12l^2p^2}}{6}.\label{eq:EHdKNAdS}\
	\end{eqnarray}
	The mass function can be written as
	\begin{equation}
		m = r_+\left(1+\frac{a^2}{l^2}+\frac{2r_+^2}{l^2}\right).\
	\end{equation}
	Using the Kerr/CFT correspondence, they have found that the central charge and CFT temperature are given by 
	\begin{eqnarray}
		c_L = \frac{12ar_+}{1+ \frac{a^2}{l^2}+\frac{6r_+^2}{l^2} },~~~~ T_L = \frac{(r_+^2 +a^2)\left(1+ \frac{a^2}{l^2}+\frac{6r_+^2}{l^2} \right)}{4\pi ar_+ \Xi}.\
	\end{eqnarray}
	The central charge and temperature above reproduce exactly the Bekenstein-Hawking entropy of extremal dyonic Kerr-Newman-AdS black hole with the help of Cardy formula.
	
	The next case is the ultraspinning version of dyonic Kerr-Newman-AdS black hole where the ultraspinning limit is still given by $a \rightarrow l$ that yields the superentropic black hole. The ultraspinning version possesses the following thermodynamic volume and conjugate of the chemical potential
	\begin{equation}
		V=\frac{2\mu}{3}r_+(r_+^2 +l^2), ~~~ K=\frac{(l^2-r_+^2)\left[(r_+^2 +l^2)^2+l^2(q^2+p^2)\right]}{8\pi r_+(r_+^2 +l^2)}. \label{eq:VKKNAdS}\
	\end{equation}
	We can write the mass and event horizon as
	\begin{equation}
		m= 2r_+ \left(\frac{r_+^2}{l^2}+1 \right), ~~~~~ r_+^2 = \frac{-l^2+l\sqrt{4l^2+3q^2+3p^2}}{3}.
	\end{equation}
	%
	Superentropic black holes violates the reverse isoperimetric inequality (RII) \cite{HennigarMannPRL2015}, which asserts
	\begin{equation}
		\mathcal{R} \equiv \left(\frac{(D-1)V}{\omega_{D-2}} \right)^{\frac{1}{D-1}}\left(\frac{\omega_{D-2}}{A} \right)^{\frac{1}{D-1}} \ge 1,
	\end{equation}
	where $A$ is the horizon area, $D$ is the dimension, and
	\begin{equation}
		\omega_D = \frac{\mu \pi^{\frac{D-1}{2}}}{\Gamma\left(\frac{D+1}{2} \right)}.\
	\end{equation}
	For superentropic dyonic Kerr-Newman-AdS black holes, we have
	\begin{equation}
		\mathcal{R} = \left(\frac{r_+^2}{r_+^2+l^2} \right)^{\frac{1}{6}}.
	\end{equation}
Since $\mathcal{R}<1$, we know that this black hole is superentropic. Using the Kerr/CFT method, for superentropic dyonic Kerr-Newman-AdS black holes, it is found that
	\begin{eqnarray}
		c_L = \frac{3\mu r_+ l^3}{\pi (3r_+^2+l^2)}, ~~~~
		T_L = \frac{(r_+^2 +l^2)(3r_+^2 +l^2)}{2\pi r_+ l^3},\
	\end{eqnarray}
which reproduces the Bekenstein-Hawking entropy for extremal ultraspinnning dyonic Kerr-Newman-AdS black hole by employing Cardy formula.
	
\subsection{Comparison}
In the previous subsection, we have considered the dyonic Kerr-Newman-AdS black hole solution and its thermodynamic quantities. The results for ultraspinning version from Kerr/CFT correspondence are also obtained. We provide the comparison of central charge, temperature, and entropy from CFT of both black holes in EMDA and Einstein-Maxwell theories in Table \ref{tab:table1}. It is clear that the main difference is the presence of dilaton and axion charges in these quantities. It is important to recall that $r_+$ is also different for both black holes. The existence of dilaton and axion charge is important in the observation of the black holes, for example to probe the existence of beyond Einstein-Maxwell theory, in this case is EMDA theory. It is crucial that for extremal dyonic Kerr-Sen-AdS black hole, $m>0, r_+^2>0, x_1^4 -16x_2^4+2x_3^4>0$. These conditions lead to $ (6r_+^2+a^2+l^2)/2 > (2r_+^2+a^2+l^2)/2 > d^2+k^2$ implying that $c_L$ is always positive. Aside from the central charge, parameters $d,k$ also define the value of the left-moving temperature and entropy as we can see in Table \ref{tab:table1}. With the same conditions as $c_L$, it can be shown that these quantities are always positive as well, similar to dyonic Kerr-Newman-AdS solution. Nonetheless, since $d,k \sim 1/m$ in dyonic Kerr-Sen-AdS black hole, it becomes the main difference from the dyonic Kerr-Newman-AdS solution because this leads to possibly more than two branches of mass with its own dual CFT.

\begin{table}[h!]
	\centering
	\caption{Comparison of CFT Quantities of Extremal Spinning Dyonic Black Holes.}
	\label{tab:table1}
	\begin{tabular}{c c c c c c c} 
		\hline
		\hline
		& Quantity &  & Dyonic Kerr-Sen-AdS & & Dyonic Kerr-Newman-AdS& \\ 
		\hline
	 \hline \\
		& $c_L$ & & {\Large $\frac{12ar_+}{1 + \frac{6r_+^2-2d^2-2k^2+a^2}{l^2}}$} & & {\Large $\frac{12ar_+}{1+\frac{6r_+^2+a^2}{l^2}} $} & \\
		\\
		\hline\\
		&  $T_L$ & &  {\Large $\frac{\left(1 + \frac{6r_+^2-2d^2-2k^2+a^2}{l^2}\right)(r_+^2-d^2-k^2+a^2)}{4\pi a r_+ \Xi}$} & &  {\Large $\frac{\left(1+\frac{6r_+^2+a^2}{l^2} \right)(r_+^2 +a^2)}{4\pi ar_+ \Xi}$} &\\
		\\
		\hline \\
		&  $S_{CFT}$ & & $\frac{\pi}{\Xi}(r_+^2 -d^2 -k^2 +a^2)$ & & $\frac{\pi}{\Xi}(r_+^2 +a^2)$ &\\
		\\
		\hline
		\hline
	\end{tabular}
\end{table}

As explained in \cite{HennigarMannPRL2015}, ultraspinning black hole might produce superentropic black hole. It means that black holes has maximum upper entropy in superentropic circumstance. We have shown that ultraspinning dyonic Kerr-Newman-AdS black hole always violates RII, likewise the Kerr-Newman-AdS black hole. However, it is different from the ultraspinning dyonic Kerr-Sen-AdS black hole where the violation depends on the value of electromagnetic charges (or dilaton and axion charges), mass and AdS length. The RII is given by
	\begin{equation}
	\mathcal{R} = \left(\frac{r_+^2}{r_+^2-d^2-k^2+l^2} \right)^{\frac{1}{6}}. \label{eq:RIIKS}
\end{equation}
\begin{table}[h!]
	\centering
	\caption{Comparison of CFT Quantities of Extremal Ultraspinning Dyonic Black Holes}
	\label{tab:table2}
	\begin{tabular}{c c c c c c c} 
		\hline
		\hline
		& Quantity &  & Dyonic Kerr-Sen-AdS & & Dyonic Kerr-Newman-AdS& \\ 
		\hline
		\hline \\
		& $c_L$ & & {\Large $\frac{3\mu r_+ l^3}{\pi\left(3r_+^2-d^2-k^2+l^2\right)} $} & & {\Large $\frac{3\mu r_+ l^3}{\pi (3r_+^2+l^2)}$} \\
		\\
		\hline\\
		&  $T_L$ & &  {\Large $\frac{\left(3r_+^2-d^2-k^2+l^2\right)(r_+^2-d^2-k^2+l^2)}{2\pi r_+ l^3 }$} & &  {\Large $\frac{(r_+^2 +l^2)(3r_+^2 +l^2)}{2\pi r_+ l^3}$} &\\
		\\
		\hline \\
		&  $S_{CFT}$ & & $\frac{\mu}{2}(r_+^2 -d^2 -k^2 +l^2)$ & & $\frac{\mu}{2}(r_+^2 +l^2)$ &\\
		\\
		\hline
		\hline
	\end{tabular}
\end{table}

Explicitly, if $0\leq q^2 +p^2 <2ml$ or $0\leq d^2 +k^2 <l^2$, RII will be violated, then the black hole is superentropic. Otherwise, if $q^2 +p^2 \geq 2ml$ or $d^2 +k^2 \geq l^2$, RII will not be violated, and the black hole is subentropic. Notably, the extremal and non-extremal ultraspinning Kerr-Sen-AdS black hole~\cite{WuWuWuYuPRD2021} are not always superentropic depending on the value of $d,k$. This is different from extremal ultraspinning dyonic Kerr-Newman-AdS black hole and its non-dyonic counterpart which are always superentropic.

However, the similarity between the ultraspinning dyonic black holes happens to be on the positivity of the CFT quantities. Similar to the Kerr-Sen-AdS black hole with general spin, we can observe from Table \ref{tab:table2} that the central charge for ultraspinning dyonic Kerr-Sen-AdS is always positive. Since from the positivity of the mass we have $r_+^2+l^2 > d^2+k^2$, then the denominator of $c_L$ will always be positive because $3r_+^2+l^2 >r_+^2+l^2 > d^2+k^2$. From Table \ref{tab:table2}, since $3r_+^2+l^2 >r_+^2+l^2 > d^2+k^2$, we can conclude that the central charge, temperature and entropy of extremal ultraspinning dyonic Kerr-Sen-AdS solution are always positive, similar to the extremal ultraspinning dyonic Kerr-Newman-AdS solution.

There is an interesting fact that central charge exactly can be negative. However, it is not for this asymptotically AdS black hole family. Yet, when we change the solution to have positive cosmological constant (or asymptotically dS) by $l^2 \rightarrow -l^2$, $c_L$ can be negative. For this `dS' type, note that we have the following relations
\begin{equation}
	m= -\frac{r_+}{l^2}(2r_+^2-2d^2-2k^2+a^2-l^2), ~~~
	r_+^2 = \frac{2d^2 +2k^2-a^2+l^2 + \sqrt{x_1^4 -16x_2^4+2x_3^4}}{6},\label{eq:EHdKSdS}\
\end{equation}
where $x_1^4=a^4 +16d^4+16k^4+l^4$, $x_2^4=a^2 d^2+a^2 k^2-d^2 l^2-k^2l^2-2d^2k^2$, $x_3^4 = -7a^2l^2 - 6l^2q^2 - 6l^2p^2$, $m>0$, and $r_+^2>0$. Negative $c_L$ can be obtained when $(6r_+^2 +a^2 -l^2)/2> d^2 + k^2 $. This is always valid as long as black hole exists, i.e., the quantity $x_1^4 -16x_2^4+2x_3^4 > 0$~(when the square root quantity is zero, exceptionally $c_{L}\to \infty$).  It is known that negative central charge exists in non-unitary CFT which may appear in ghost system from string theory \cite{NarayanPRD}. Moreover, there is possibly more than two branches of mass for this black hole resulting in more than two branches of CFT with different central charges just like in the AdS type. Notably, when we consider RII for $l^2 \rightarrow -l^2$ in (\ref{eq:RIIKS}) it is obvious that $\mathcal{R}>1$ implying that superentropic black hole cannot be obtained.

\section{Conclusion}
\label{eq:conclusion}
We have demonstrated that the Kerr/CFT correspondence is explicitly well-defined for a dyonic black hole family in EMDA theory, i. e. dyonic Kerr-Sen black hole, its gauged and ultraspinning counterparts. Our work enlarges the family of metrics respecting Kerr/CFT correspondence for dyonic black holes in EMDA theory. For dyonic Kerr-sen black hole, we have obtained that the central charge is identical to Kerr-Sen and Kerr black holes, yet with different constraint on the mass parameter due to the presence of dilaton and axion charges. Interestingly, due to the presence of these charges, we have found that the mass possesses two different branches leading also to the central charge of the CFTs with two different branches. The main upshot to prove that the entropy from CFT matches with Bekenstein-Hawking entropy has succeeded for this dyonic black hole family in EMDA theory.

We then extend the duality calculation to the gauged and ultraspinning counterparts. Since there exist more than two horizons, we have had to approximate the single event horizon for its extremal case. Since for non-gauged dyonic Kerr-Sen black hole the central charge could have two branches, we have argued that possibly there should be more than two branches of the mass for gauged case that leads to more than two branches of the central charge and entropy. We have found that finite central charges for both gauged and ultraspinning black hole solutions reproduce Bekenstein-Hawking entropy. It has been proven also that the central charge, temperature and entropy for these black holes are always positive. This extremal ultraspinning dyonic black hole and its non-extremal counterpart cannot always be superentropic in the ultraspinning limit depending on the value of $d,k$. We have also considered the asymptotically dS solution with $l^2 \rightarrow -l^2$ substitution. Interestingly in this case, the central charge can be negative unlike the AdS type. It is well-known that negative central charge might appear in non-unitary CFT, for example in string theory. It will be interesting for future investigation.

In conclusion, our calculations have supported the Kerr/CFT correspondence conjecture. This result suggests that the extremal rotating dyonic black holes in EMDA theory is holographically dual to 2D CFT represented by non-vanishing left-moving central charge. For the future investigation, it would be intriguing to explore the hidden conformal symmetry from non-extremal dyonic black hole family in EMDA theory using the scalar field and higher spin fields.

\vspace*{8mm}

\noindent {\large{\bf Acknowledgments}}
\vspace*{4mm}

We would like to thank the anonymous referee for valuable comments which help improve our work considerably. M. F. A. R. S. is supported by the Second Century Fund (C2F), Chulalongkorn University, Thailand.  \newline


\vspace*{1mm}

\begin{thebibliography}{99}

\bibitem{StromingerVafaPLB1996}	
A. Strominger and C. Vafa, \textit{Microscopic origin of the
Bekenstein-Hawking entropy}, \textit{Phys. Lett.} \textbf{B 379} (1996) 99.

\bibitem{SenGRG2008}
A. Sen, \textit{Black hole entropy function, attractors and precision counting of microstates},\textit{ Gen. Relativ. Gravit.} \textbf{40} (2008) 2249.


\bibitem{GuicaPRD2009} 
M.~Guica, T.~Hartman, W.~Song and A.~Strominger, \textit{The Kerr/CFT Correspondence},  \textit{Phys. Rev.} {\bf D 80} (2009) 124008.

	

\bibitem{brown1986}
J. D. Brown and M. Henneaux, \textit{Central charges in the canonical realization of asymptotic symmetries: an example from three-dimensional gravity}, \textit{Commun. Math. Phys.} \textbf{104} (1986) 207.
				
\bibitem{Hartman2009}
T. Hartman, K. Murata, T. Nishioka and A. Strominger, \textit{CFT duals for extreme black holes}, \textit{JHEP} \textbf{04} (2009) 019. 
			
\bibitem{Ghezelbash2009}
A. M. Ghezelbash, \textit{Kerr/CFT correspondence in the low energy limit of heterotic string theory}, \textit{JHEP} \textbf{08} (2009) 045.
			
\bibitem{Lu2009}
H. L{\"u}, J. Mei and C. N. Pope, \textit{Kerr-AdS/CFT correspondence in diverse dimensions}, \textit{JHEP} \textbf{04} (2009) 054.
				
\bibitem{Li2010}
R. Li and J.-R. Ren, \textit{Holographic dual of linear dilaton black hole in Einstein-Maxwell-Dilaton-Axion gravity}, \textit{JHEP} \textbf{09} (2010) 039. 
				
\bibitem{Anninos2010}
D. Anninos and T. Hartman, \textit{Holography at an extremal de Sitter horizon}, \textit{JHEP} \textbf{03} (2010) 096. 
				
\bibitem{Ghodsi2010}
A. Ghodsi and M. R. Garousi, \textit{The RN/CFT correspondence} \textit{Phys. Lett.} \textbf{B 687} (2010) 79. 
					
\bibitem{Ghezelbash2012}
A. M. Ghezelbash, \textit{Kerr-Bolt spacetimes and Kerr/CFT correspondence}, \textit{Mod. Phys. Lett.} \textbf{A 27} (2012) 1250046. 
						
\bibitem{Astorino2015}
M. Astorino, \textit{Microscopic entropy of the magnetised extremal Reissner-Nordstr{\"o}m black hole}, \textit{JHEP} \textbf{10} (2015) 016. 
						
\bibitem{Astorino2015a}
M.  Astorino, \textit{Magnetised Kerr/CFT correspondence}, \textit{Phys. Lett.} \textbf{B 751} (2015) 96. 
							
\bibitem{Siahaan2016}
H. M. Siahaan, \textit{Magnetized Kerr/CFT correspondence}, \textit{Class. Quantum Grav.} \textbf{33} (2016) 155013. 
								
\bibitem{Astorino2016}
M. Astorino, \textit{CFT duals for accelerating black holes}, \textit{Phys. Lett.} \textbf{B 760} (2016) 393. 
									
\bibitem{Sinamuli2016}
M. Sinamuli and R. B. Mann, \textit{Super-entropic black holes and the Kerr-CFT correspondence}, \textit{JHEP} \textbf{08} (2016) 148. 
									
\bibitem{Sakti2018}
M. F. A. R. Sakti, A. Suroso and F. P. Zen, \textit{CFT duals on extremal rotating NUT black holes}, \textit{Int. J. Mod. Phys.}  \textbf{D 27} (2018) 1850109. 
										
\bibitem{SaktiEPJPlus2019}
M. F. A. R. Sakti, A. Suroso and F. P. Zen, \textit{Kerr/CFT correspondence on Kerr-Newman-NUT-Quintessence black hole}, \textit{Eur. Phys. J. Plus} \textbf{134} (2019) 580. 
											
\bibitem{SaktiMicroJPCS2019}
M. F. A. R. Sakti, A. Suroso and F. P. Zen, \textit{Microscopic entropy of extremal Kerr black holes}, \textit{J. Phys.: Conf. Ser.} \textbf{1204} (2019) 012009. 

\bibitem{SaktiAnnPhys2020}
M. F. A. R. Sakti, A. Suroso and F. P. Zen, \textit{Kerr-Newman-NUT-Kiselev black holes in Rastall theory of gravity and Kerr/CFT correspondence}, \textit{Ann. Phys.} \textbf{413} (2020) 168062. 
	
\bibitem{SaktiPhysDarkU2021}
M. F. A. R. Sakti and F. P. Zen, \textit{CFT duals on rotating charged black holes surrounded by quintessence}, \textit{Phys. Dark Universe} \textbf{31} (2021) 100778. 
												
\bibitem{Castro2010}
A. Castro, A. Maloney and A. Strominger, \textit{Hidden conformal symmetry of the Kerr black hole}, \textit{Phys. Rev.} \textbf{D 82} (2010) 024008. 

\bibitem{LoweSkanataPRD2014}
D. A. Lowe, Antun Skanata, and I. Messamah, \textit{Hidden Kerr/CFT correspondence at finite frequencies}, \textit{Phys. Rev.} \textbf{D 89} (2014) 064005.
												
\bibitem{ChenLong2010}
B. Chen, J. Long and J.-J. Zhang, \textit{Hidden conformal symmetry of extremal black holes}, \textit{Phys. Rev.} \textbf{D 82} (2010) 104017. 

\bibitem{CveticLarsenJHEP2009}
M. Cveti$\breve{\text{c}}$ and F. Larsen, \textit{Greybody factors and charges in Kerr/CFT}, \textit{JHEP} \textbf{09} (2009) 088. 
												
\bibitem{ChenSun2010}
C.-M. Chen and J.-R. Sun, \textit{Hidden conformal symmetry of the Reissner-Nordstr{\"o}m black holes}, \textit{JHEP} \textbf{08} (2010) 034. 
												
\bibitem{ChenWang2010}
D. Chen, H. Wang, H. Wu and H. Yang, \textit{Hidden conformal symmetry of extreme and non-extreme Einstein-Maxwell-Dilaton-Axion black holes}, \textit{JHEP} \textbf{11} (2010) 002. 
												
\bibitem{WangLiu2010}
Y.-Q. Wang and Y.-X. Liu, \textit{Hidden conformal symmetry of the Kerr-Newman black hole}, \textit{JHEP} \textbf{08} (2010) 087. 
												
\bibitem{ChenLongJHEP2010}
B. Chen and J. Long, \textit{On holographic description of the Kerr-Newman-AdS-dS black holes}, \textit{JHEP} \textbf{08} (2010) 065. 
												
\bibitem{SetareKamali2010}
M. R. Setare and V. Kamali, \textit{Hidden conformal symmetry of extremal Kerr-Bolt spacetimes}, \textit{JHEP} \textbf{10} (2010) 074. 
												
\bibitem{GhezelbashKamali2010}
A. M. Ghezelbash, V. Kamali and M. R. Setare, \textit{Hidden conformal symmetry of Kerr-Bolt spacetimes}, \textit{Phys. Rev.} \textbf{D 82} (2010) 124051. 
													
\bibitem{ChenHuangPRD2010}
C.-M. Chen, Y.-M. Huang, J.-R. Sun, M.-F. Wu and S.-J. Zou, \textit{Twofold hidden conformal symmetries of the Kerr-Newman black hole}, \textit{Phys. Rev.} \textbf{D 82} (2010) 066004. 
													
\bibitem{ChenChen2011}
B. Chen, C.-M. Chen and B. Ning, \textit{Holographic Q-picture of Kerr-Newman-AdS-dS black hole}, \textit{Nuc. Phys.} \textbf{B 853} (2011) 196-209.

														
\bibitem{ChenGhezelbash2011}
B. Chen, A. M. Ghezelbash, V. Kamali and M. R. Setare, \textit{Holographic description of Kerr-Bolt-AdS-dS spacetimes}, \textit{Nuc. Phys.} \textbf{B 848} (2011) 108-120. 
															
\bibitem{ChenZhangJHEP2011}
B. Chen and J.-J. Zhang, \textit{General hidden conformal symmetry of 4D Kerr-Newman and 5D Kerr black holes}, \textit{JHEP} \textbf{08} (2011) 114. 
															
\bibitem{HuangYuan2011}
Y.-C. Huang and F.-F. Yuan, \textit{Hidden conformal symmetry of extremal Kaluza-Klein black hole in four dimensions}, \textit{JHEP} \textbf{03} (2011) 029. 
															
\bibitem{Shao2011}
K.-N. Shao and Z. Zhang, \textit{Hidden conformal symmetry of a rotating black hole with four charges}, \textit{Phys. Rev.} \textbf{D 83} (2011) 106008. 
															
\bibitem{DeyouChen2011}
D. Chen, P. Wang, H. Wu and H. Yang, \textit{Hidden conformal symmetry of rotating charged black holes}, \textit{Gen. Relativ. Grav.} \textbf{43} (2011) 181-190. 
		
\bibitem{CveticLarsenJHEP2012a}
M. Cveti$\breve{\text{c}}$ and F. Larsen, \textit{Conformal symmetry for general black holes}, \textit{JHEP} \textbf{02} (2012) 122.

\bibitem{CveticLarsenJHEP2012b}
M. Cveti$\breve{\text{c}}$ and F. Larsen, \textit{Conformal symmetry for black holes in four dimensions}, \textit{JHEP} \textbf{09} (2012) 076.
														
\bibitem{GhezelbashSiahaanCQG2013}
A. M. Ghezelbash and H. M. Siahaan, \textit{Hidden and generalized conformal symmetry of Kerr–Sen spacetimes}, \textit{Class. Quantum Grav.} \textbf{30} (2013) 135005.

\bibitem{GhezelbashSiahaanGRG2014}
A. M. Ghezelbash and H. M. Siahaan, \textit{Deformed hidden conformal symmetry for rotating black holes}, \textit{Gen. Relativ. Grav.} \textbf{46} (2014) 1783. 
																
\bibitem{SiahaanAcc2018}
H. M. Siahaan, \textit{Hidden conformal symmetry for the accelerating Kerr black holes}, \textit{Class. Quantum Grav.} \textbf{35} (2018) 155002. 
	
\bibitem{Saktideformed2019}
M. F. A. R. Sakti, A. M. Ghezelbash, A. Suroso and F. P. Zen, \textit{Deformed conformal symmetry of Kerr-Newman-NUT-AdS black holes}, \textit{Gen. Relativ. Grav.} \textbf{51} (2019) 151. 

\bibitem{SaktiNucPhysB2020}
M. F. A. R. Sakti, A. M. Ghezelbash, A. Suroso and F. P. Zen,\textit{ Hidden conformal symmetry for Kerr-Newman-NUT-AdS black holes}, \textit{Nuc. Phys.} \textbf{B 953} (2020) 114970. 

\bibitem{SaktiPhysDarkU2022}
M. F. A. R. Sakti, A. Suroso, A. Sulaksono and F. P. Zen, \textit{Rotating black holes and exotic compact objects in the Kerr/CFT correspondence within Rastall gravity}, \textit{Phys. Dark Universe} \textbf{35} (2022) 100974. 

\bibitem{Saktidyonicnonex}
M. F. A. R. Sakti, \textit{Hidden conformal symmetry for dyonic Kerr-Sen black hole and its gauged family}, arXiv:2208.02722 [hep-th].

\bibitem{HacoJHEP2018}
S. Haco, S. W. Hawking, M. J. Perry, and A. Strominger, \textit{Black hole entropy and soft hair}, \textit{JHEP} \textbf{12} (2018) 098. 

\bibitem{AggarwalCastroJHEP2020}
A. Aggarwal, A. Castro and S. Detournay, \textit{Warped symmetries of the Kerr black hole}, \textit{JHEP} \textbf{01} (2020) 016.

\bibitem{WuWuWuYuPRD2021}
D. Wu, S.-Q. Wu, P. Wu, and H. Yu, \textit{Aspects of the dyonic Kerr-sen-AdS$_4$ black hole and its ultraspinning version}, \textit{Phys. Rev.} \textbf{D 103} (2021) 044014.

\bibitem{WuWuYuWuPRD2020}
D. Wu, P. Wu, H. Yu, and S.-Q. Wu, \textit{Are ultraspinning KerrSen-AdS4 black holes always super-entropic?}, \textit{Phys. Rev.} \textbf{D 102} (2020) 044007.

\bibitem{HennigarMannPRL2015}
R. A. Hennigar, R. B. Mann, and D. Kubizň{\' a}k, \textit{Entropy
Inequality Violations from Ultraspinning Black Holes},
\textit{Phys. Rev. Lett.} \textbf{115} (2015) 031101.

\bibitem{GaltsovPRL1995}
A. Garc{\'i}a, D. Galtsov, and O. Kechkin, \textit{Class of Stationary Axisymmetric Solutions of the Einstein-Maxwell-Dilaton-Axion Field Equations}, \textit{Phys. Rev. Lett.} \textbf{74} (1995) 1276.

\bibitem{ChongCveticNPB2005}
 Z. W. Chong, M. Cveti$\breve{\text{c}}$, H. Lü, and C. N. Pope, \textit{Charged rotating black holes in four-dimensional gauged and ungauged supergravities}, \textit{Nucl. Phys.} \textbf{B 717} (2005) 246.

\bibitem{ChowCompereCQG2014}
D. D. K. Chow and G. Comp{\'e}re, \textit{Seed for general rotating non-extremal black holes of $N = 8$ supergravity}, \textit{Class. Quantum Grav.} \textbf{31} (2014) 022001.

\bibitem{ChowComperePRD2014}D. D. K. Chow and G. Comp{\'e}re, \textit{Black holes in $\mathcal{N}$=  8 supergravity from SO(4,4) hidden symmetries}, \textit{Phys. Rev.} \textbf{90} (2014) 025029.
								
\bibitem{Compere2017}
G. Comp{\'e}re, \textit{The Kerr/CFT correspondence and its extensions}, \textit{Living Rev. Rel.} \textbf{15} (2012) 11 [\textit{Addendum ibid.} \textbf{20} (2017) 1]. 
																					


\bibitem{CveticPopePRD2011}
M. Cveti$\breve{\text{c}}$, G. W. Gibbons, D. Kubiz$\breve{\text{n}}${\'a}k, and C. N. Pope, \textit{Black hole enthalpy and an entropy inequality for the thermodynamic volume}, \textit{Phys. Rev.} \textbf{D 84} (2011) 024037.

\bibitem{BarnichBrandt2002}
G. Barnich and F. Brandt, \textit{Covariant theory of asymptotic symmetries, conservation laws and central charges}, \textit{Nuc. Phys.} \textbf{B 633} (2002) 3. 

\bibitem{NarayanPRD}
K. Narayan, \textit{On dS$_4$ extremal surfaces and entanglement entropy in some ghost CFTs}, \textit{Phys. Rev.} \textbf{D 94} (2016) 046001.

\bibitem{HerdeiroPLB2021}
C. Herdeiro, E. Radu, and K.Uzawa, \textit{De-singularizing the extremal GMGHS black hole via higher derivatives corrections}, \textit{Phys. Lett.} \textbf{B 818} (2021) 136357.

\end{thebibliography}
\end{document}